\documentclass[sigconf, screen]{acmart}
\PassOptionsToPackage{dvipsnames}{xcolor}

\PassOptionsToPackage{colorlinks=true,linkcolor=RubineRed,breaklinks=True,citecolor=RubineRed}{xcolor}
\usepackage[normalem]{ulem}
\usepackage{framed}

\usepackage{enumitem}
\setlist[itemize]{noitemsep, topsep=0pt}

\usepackage[most]{tcolorbox}
\usepackage{adjustbox}
\usepackage[linesnumbered,algo2e,ruled]{algorithm2e}
\usepackage{multirow}
\usepackage{booktabs,subcaption,dcolumn}
\usepackage{tikz}
\usepackage{enumitem}
\usepackage{tabularx}
\usepackage{MnSymbol}

\usepackage{pifont}
\usepackage{svg}
\usepackage{soul}
\usepackage{amsmath}
\usepackage{dirtytalk}
\usepackage{subcaption}
\usepackage{graphicx}
\usepackage{setspace}
\usepackage[font=small]{caption}
\usepackage{booktabs} 
\usepackage[titletoc,toc,title]{appendix}
\usepackage{svg}
\usepackage{amsfonts}
\usepackage{xurl} 

\usepackage{diagbox}
\usepackage{indentfirst}

\usepackage{CJKutf8}
\usepackage{amsmath}
\usepackage{amsfonts} 

\usepackage{colortbl}

\usepackage{siunitx}
\usepackage[table]{xcolor}
\usepackage{setspace}

\AtBeginDocument{%
  \providecommand\BibTeX{{%
    \normalfont B\kern-0.5em{\scshape i\kern-0.25em b}\kern-0.8em\TeX}}}

\acmYear{2026}\copyrightyear{2026}
\setcopyright{cc}
\setcctype[4.0]{by-nc-nd}
\acmConference[ASIA CCS '26]{ACM Asia Conference on Computer and Communications Security}{June 1--5, 2026}{Bangalore, India}
\acmBooktitle{ACM Asia Conference on Computer and Communications Security (ASIA CCS '26), June 1--5, 2026, Bangalore, India}
\acmDOI{10.1145/3779208.3807488}
\acmISBN{979-8-4007-2356-8/26/06}

\newcolumntype{?}{!{\vrule width 1.5pt}}

\newcommand{\textbox}[1]{
    \noindent\fbox{%
        \parbox{0.97\columnwidth}{%
            {#1}
        }%
    }
}

\newcommand\smamath[1]{{\small $#1$}}

\newcommand\scmath[1]{{\scriptsize $#1$}}

\newcommand\revision[1]{%
  \bgroup
  \hskip0pt\color{blue!80!black}%
  #1%
  \egroup
}

\input{auxiliary/acm_variant}

\begin{document}
\title[SoK: Reshaping Research on Network Intrusion Detection Systems]{SoK: Reshaping Research on\\Network Intrusion Detection Systems}
\author{Giovanni Apruzzese}
\orcid{0000-0002-6890-9611}
\affiliation{%
  \institution{University of Liechtenstein}
  \city{Vaduz}
  \country{Liechtenstein}}
\affiliation{%
  \institution{Reykjavik University}
  \city{Reykjavik}
  \country{Iceland}}
\email{giovannia@ru.is}


\begin{abstract}
Network Intrusion Detection Systems (NIDS) have been studied for decades. Hundreds of papers have, e.g., proposed ways to enhance, harden or bypass NIDS. However, the findings of prior literature are hardly reflected in real-world operational contexts. Such a disconnection is problematic for research itself: it is unclear what scenario envisioned by prior work can be used as a baseline for future advancements.

We argue that a key reason for this disconnection is a fundamental misunderstanding of intrinsic characteristics of NIDS.
For instance, the fact that a compromised NIDS cannot be expected to work well; the fact that some evaluations are done without carrying out any experiment in a (even synthetic) ``real'' network; the fact that security operators triage high-level reports---and not individual samples flagged by some classifier. In this SoK, which is primarily a reflective piece, we first constructively highlight such quintessential properties (without criticizing \textit{any} work by different authors) by stating three Assertions.
Then, we provide recommendations---further emphasized through an original and reproducible case study that challenges some established practices. Ultimately, we seek to lay a foundation to reshape research on NIDS.
\end{abstract}



\settopmatter{printfolios=true}

\maketitle

\section{Introduction}
\label{sec:introduction}

\begin{figure}[t]
    \centering
    \includegraphics[width=0.99\columnwidth]{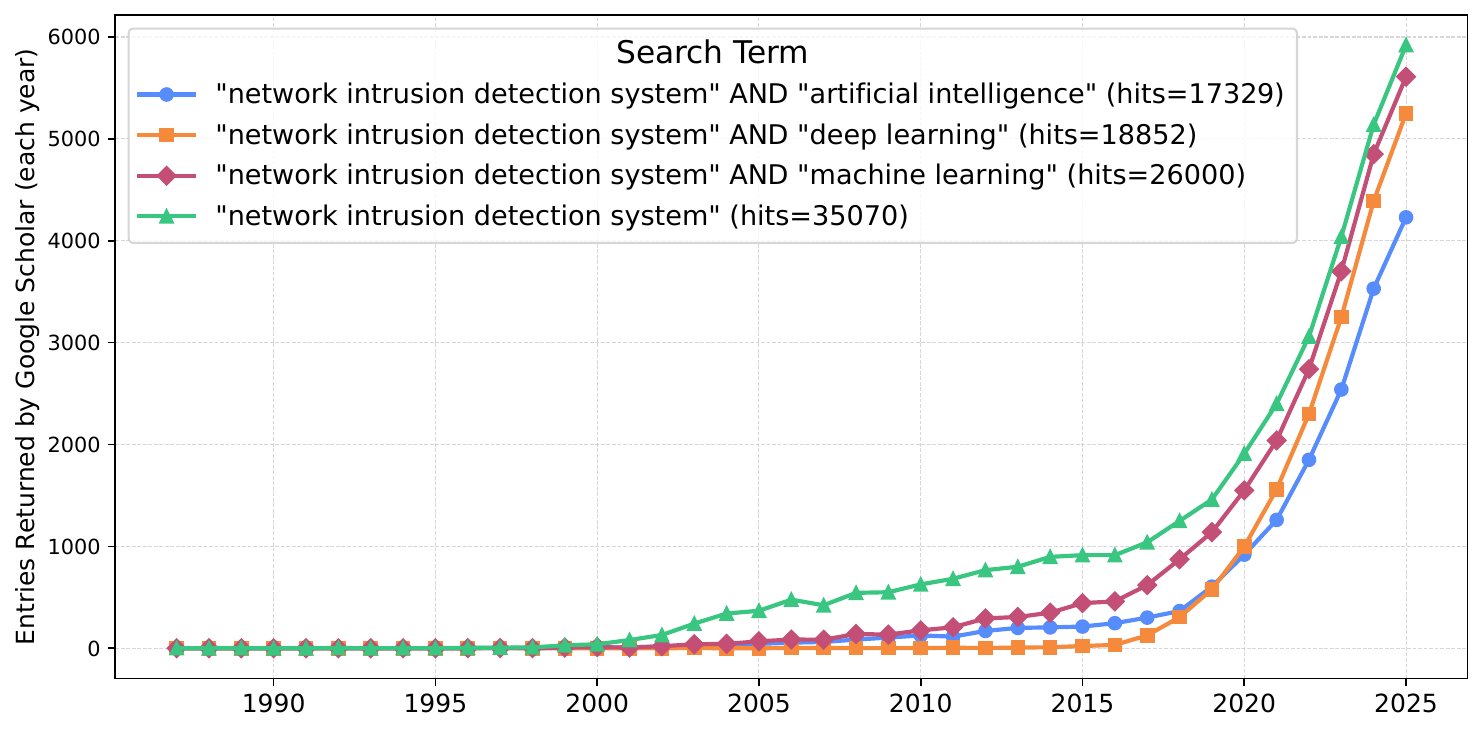}
    \vspace{-4mm}
    \caption{\textbf{Research interest in NIDS.}
    \textmd{Queries issued on March 2026.}
    } 
    \label{fig:scholar}
    \vspace{-3mm}
\end{figure}

\noindent
Research on Network Intrusion Detection Systems (NIDS) is constantly increasing (see Fig.~\ref{fig:scholar}). Since 2015, thousands of articles appear yearly on NIDS, especially in conjunction with machine-learning (ML) techniques. Such articles propose new methods that improve NIDS-related components~\cite{andresini2021insomnia}, or extend NIDS functionalities~\cite{yang2021cade}, or bypass existing NIDS~\cite{ayub2020model}; some papers also systematized~\cite{apruzzese2023sok} or ``troubleshooted''~\cite{engelen2021troubleshooting} our knowledge on this domain.

We acknowledge the contributions of prior work. However, we believe that, over time, scientific literature on NIDS has been focusing excessively on the ``research'' aspect---losing sight of the real-world implications of NIDS-related technologies. As a potential consequence, a recent survey among industry practitioners revealed skepticism towards the results claimed in research~\cite{apruzzese2023sok}.

We want to change this.
This ``systematization of knowledge'' (SoK) paper attempts to trigger a reflective exercise in the reader. We  do so by stating three \textsc{assertions}, rooted in established security principles, and which we believe should not be forgotten by future work aiming to advance the state of the art in NIDS.

\subsection{Goal and Methodology (and disclaimer)}
\label{ssec:goal}

\noindent
This SoK paper is rooted on the following assumption: \textit{research in NIDS is relatively stagnant, and is hindered by, among others, a superficial treatment of the NIDS domain}. By ``superficial treatment,'' we intend: {\small \textit{(a)}}~overlooking essential security principles; or {\small \textit{(b)}}~ignoring the characteristics of modern networks; or {\small \textit{(c)}}~neglecting the basic function of NIDS---or any combination of the previous points.\footnote{The assumptions/\textsc{Assertions} of this paper are not only supported by this paper's authors, but also by fellow (anonymous) researchers who reviewed previous drafts (provided in~\cite{repository}). Yet, while it is objectively impossible to ``universally confirm'' this paper's overarching assumption, we provide supporting evidence in Section~§\ref{ssec:motivation}.}

Hence, our goal is to provide a constructive foundation that can put an end to this stagnation. Specifically, our overarching objective is to devise a concise \textit{vademecum}---a guiding reference---that can help (re)shape future research on NIDS. In doing so, we adopt a positive stance toward prior literature. Indeed, we will treat past works by following three rules:
\begin{itemize}[leftmargin=7mm, topsep=0pt]
    \item[1)] We will use prior work to provide a basis for the arguments and \textsc{assertions} stated in this paper.
    \item[2)] We will use prior work to suggest avenues for future work, or highlight exemplary good practices. 
    \item[3)] We will never explicitly criticize any prior work---barred those co-authored by this paper's co-authors.
\end{itemize}
In complying with the first two rules, we will draw from both recent literature, published in so-called ``high-ranked venues'' (such as~\cite{flood2024bad}, best paper award at EuroS\&P'24); as well as from seminal works from the previous century---including some published when the concept of ``NIDS'' had not yet been conceived (such as the 1984 paper by Thompson on ``Reflections on trusting trust''~\cite{thompson1984reflections}). 

Moreover, by adhering to the third rule, we will not attempt to ``point the finger at'' works which may not follow any of the assertions stated in this paper---except for those co-authored by the authors of this piece (such as~\cite{apruzzese2020deep}). Therefore, readers who expect to, e.g., find explicit mention of papers that ``violate'' any given assertion---even with the (legitimate) goal to motivate that there is a need for such an assertion---will be disappointed. We firmly believe that doing so is not constructive.

\subsection{Target Audience and Contributions}
\label{ssec:audience}

\noindent
This paper is addressed to any researcher with an interest in NIDS. We identify two classes of potential readers:
\begin{itemize}[leftmargin=*, topsep=0pt]
    \item People who want to carry out research in NIDS. This includes both the act of \textit{designing} a plan to answer a given research question; or \textit{writing} a document that summarizes or advances the state of the art; but also \textit{reading} prior related work.
    
    \item People who serve as reviewers of NIDS-related documents. This includes both those who must gauge the scientific value of a given document (e.g., to determine whether the document should be ``accepted'' or ``rejected'' in a peer-reviewed selection procedure) and those who are solicited to provide feedback and constructive criticism on such a document to improve its quality.
\end{itemize}
Hence, this paper is meant to be understood by those with long-term expertise in NIDS, or by those who have tackled NIDS in orthogonal domains (e.g., applied ML, which is closely tied to NIDS as shown by Fig.~\ref{fig:scholar}), as well as by those who are just approaching any NIDS-related field. Therefore, we will favor simple logical arguments, rooted on elementary security principles, and we will deliberately not include complex notation or domain-specific jargon.

The contributions of this work are organized as follows. 
First, in Section~§\ref{sec:2}, we define our scope and provide the essential terminology, through which we also clarify potential misconceptions.
Then, in Sections~§\ref{sec:assertion_1} to~§\ref{sec:assertion_3}, we state our \textbf{three assertions}---each focusing on a specific ``issue'' of NIDS-related research. We make some considerations that justify each assertion, and discuss implications and recommendations for future work. 
Next, in Section~§\ref{sec:demonstration}, we provide an exemplary demonstration of how our assertions can be used in NIDS-related research: we will do so with a \textbf{novel experiment}, through which we uncover results that substantially question prior knowledge in the field (our code is open source~\cite{repository}).
Lastly, in Section~§\ref{sec:discussion}, we reflect on our assertions, clarifying potential misunderstandings, and also derive our \textbf{vademecum} (in Table~\ref{tab:vademecum}).
Conclusions are drawn in Section~§\ref{sec:conclusions}. The Appendix complements this piece with a glossary of NIDS (taken from the RFC~\cite{shirey2007internet}), and other real-world evidence supporting this paper's arguments.

\section{Context and Preliminaries}
\label{sec:2}

\noindent
To provide a constructive systematization that addresses the problem tackled by this paper, i.e., the ``superficial treatment of the NIDS domain'', we first define the scope of our work (§\ref{ssec:scope}). Then, we define some key terms that can be source of misunderstandings (§\ref{ssec:terminology}). Finally, we conclude by describing a use-case of a NIDS (§\ref{ssec:scenario}).


\subsection{Scope: \textit{Network} Intrusion Detection Systems}
\label{ssec:scope}

\noindent
We focus on Network Intrusion Detection Systems, i.e., NIDS. This term is strongly related to the broader ``Intrusion Detection System'' (IDS), which has been extensively discussed in the last 38 years (e.g.,~\cite{biermann2001comparison,khraisat2019survey,vasilomanolakis2015taxonomy,apruzzese2023sok}), starting from the seminal work by Denning~\cite{denning1987intrusion} in 1987. 

The distinction between IDS and NIDS is the term ``network'', which can be understood in various ways. Among the most common interpretations, ``network'' can refer to the focus (or level) of the analysis---e.g., an IDS whose goal is to detect intrusion in network of hosts (in contrast to, e.g., an IDS which focuses on one host~\cite{bace2001intrusion}); or to the data used to perform the detection---e.g., an IDS that analyzes network data~\cite{debar1999towards} (in contrast to, e.g., host-level logs). 

Here, we refer to an NIDS as an IDS that fulfills both of the aforementioned criteria. This is to define a clear separation with other types of IDS, such as those that focus on detecting intrusions at the host-level (e.g., HIDS) by using logs~\cite{frasao2024see}, source-code analysis~\cite{egele2006using}, or provenance graphs (e.g.,~\cite{liu2018host,bridges2019survey,cheng2024kairos,jia2024magic,alsaheel2021atlas}); or even those that do use network-related data, but ``centered'' on a single host~\cite{liu2022collaborative,abad2003log}.\footnote{Most assertions and recommendations made in this paper apply also to these complementary classes of IDS, as well as to any security system (e.g., those devoted to malware detection~\cite{dambra2023decoding,tian2025density,alasmary2019analyzing}, or explainability~\cite{he2023finer}, or protection of cyber-physical systems~\cite{grochocki2012ami,chen2012smart,erba2020constrained,luo2021deep}, as well as those for the automotive context~\cite{pollicino2024performance,longari2020cannolo,liu2025vehicular,cerracchio2024investigating}).}

\vspace{1mm}
\textbox{{We refer to a NIDS as \textit{a system that focuses on detecting intrusions in a network by analysing network-related data}.}}
\vspace{1mm}

We report in the Appendix a glossary providing the actual definitions, taken verbatim from the RFC, of technical terms such as ``system'', ``intrusion'', ``network''. Importantly, we stress that the goal of an (N)IDS is to \textit{detect intrusions}, i.e., security-noteworthy events that assume that an entity has obtained (or attempts to obtain) unauthorized access to a given system or resource. From this, we derive two important considerations:
{\small \textit{(i)}}~the focus is to ``detect'' intrusions as early and effectively as possible to minimise damage---and not to prevent intrusions;\footnote{We do acknowledge, however, that some (N)IDS may integrate some form of ``prevention'' mechanism that may automatically enact certain policies.} and
{\small \textit{(ii)}}~to ``detect an intrusion,'' it is implicit that the intrusion must have already occurred---in other words, (N)IDS operate with the assumption that attackers may have already acquired a foothold in the monitored network.

\subsection{Terminology (and clarifications)}
\label{ssec:terminology}

\noindent
We discuss some terms associated with our interpretation of NIDS, in an attempt to prevent generating some misunderstandings (which we ourselves encountered), further clarify our scope, and provide some suggestions for future work.

\textbf{Detection \textit{approaches}.} There exist various ways in which a NIDS can fulfill its detection objectives. Common terms are ``signature-based'' or ``anomaly-based'' approaches~\cite{apruzzese2023sok,yang2022systematic,lee2017athena,song2024madeline}: the former detect intrusions by trying to identify known patterns of malicious behaviour in the analysed data, whereas the latter seek to capture malicious events when there is some deviation from an established notion of normality. Both of these approaches have their pros and cons, and can be implemented either manually---by relying on heuristics, security feeds, organization-specific policies, or domain expertise; or via data-driven techniques---including those pertaining to the machine-learning (ML) domain, either supervised or unsupervised~\cite{wang2023network,Mirsky:Kitsune,yang2021cade,van2022deepcase,ceschin2024machine,Sommer:Outside,zanero2004unsupervised}, as well as reinforcement-learning based (e.g.,~\cite{apruzzese2020deep}); moreover, both signature- and anomaly-based approaches can work ``online'' (i.e., by analysing data in real time~\cite{gyamfi2022novel}) or ``offline'' (i.e., by analysing data in batches~\cite{horchulhack2022toward}). This paper covers all of these. 

\textbf{What is an \textit{anomaly}?}
We emphasize that \textit{an anomaly is not necessarily a security-noteworthy event}, and vice-versa. For instance, it is possible that a certain piece of software begins performing ``novel'' network activities which are benign in nature but are flagged as anomalies by the NIDS. Conversely, it is possible that a malicious host, in an attempt to evade an anomaly-based NIDS, performs network activities that mimic normal traffic patterns, therefore not triggering any anomaly by the NIDS. Such an observation is crucial: a NIDS that detects a lot of anomalies, even when these anomalies are ``true'' anomalies, is not necessarily a NIDS that works well from a security standpoint. In light of this, we endorse future work focusing on anomaly-based NIDS to clearly state what is the objective of the anomaly-detection mechanism.

\textbf{\textit{Evasion} of NIDS.}
The term ``evasion'' can have many meanings in the security domain. Particularly, in the adversarial ML context, it is often used to denote attacks at ``test time''~\cite{biggio2013evasion}, which can be misleading~\cite{apruzzese2023real}. In this work, we define ``evasion'' in its literal sense. Given that we are considering a scenario in which there is an NIDS that inspects network traffic to detect intrusions in the network, an ``evasion'' can be considered as any circumstance in which an ``intrusion'' is not detected by the NIDS. In other words, our definition of evasion encompasses both that of ``adversarial ML attacks'' (in which a malicious sample is classified as benign via an adversarial perturbation introduced at test time~\cite{han2021evaluating}), as well as those achieved via, e.g., concealment attacks~\cite{erba2020constrained}, and even those that are due to poor configuration of the NIDS~\cite{handley2001network}; the latter, however, cannot be considered as attacks because there is no deliberate intention by the attacker to evade the NIDS (see: RFC definitions of ``attack'' in Appendix~\ref{app:glossary}). 

\textbf{What is the \textit{output} of an NIDS?} 
A NIDS generates \textit{alerts} (or \textit{alarms})~\cite{fung2024attributions,debar2001aggregation,shah2019methodology,ban2023breaking}: if the NIDS believes to have identified a security-noteworthy event, allegedly resembling an intrusion, then the NIDS raises one (or more) alerts. Such alerts are meant to be triaged by a human operator---potentially a member of the security operation center (SOC): if the alerts were truly representative of a security incident, then appropriate mitigations would be enacted; otherwise, the alarms would be considered as ``false positives'' and no action would be taken~\cite{alahmadi202299}. Nonetheless, even though NIDS may integrate some classification mechanisms that discriminate between benign and malicious datapoints, it is after post-processing the output of the NIDS (typically via automated means~\cite{van2022deepcase}) that an actual decision will be made. Abundant efforts now focus on developing ``dashboards'' that can facilitate the decision-making of SOC-analysts~\cite{muhammad2023integrated,gonzalez2021security} by visualizing the output of NIDS in actionable reports---a function fulfilled by so-called System Information and Event Management (SIEM) tools.

\vspace{1mm}
\textbox{\textbf{A comment on ``unrealistic.''} In the security domain, it is common to encounter the term ``unrealistic'' (e.g., ``this threat model is \textit{unrealistic}''). We believe that this term is used with different meanings, which can lead to misunderstandings. In some cases, it is used to identify a scenario that is ``unlikely to occur in reality''; in other cases, to identify a scenario that is ``impossible to occur in reality''. These two expressions are semantically very different: the latter is a clear indication of impossibility, whereas the former, despite being rather vague, does not exclude that a given scenario may occur. We advise readers to refrain from using this term on its own and without providing additional context.}

\subsection{Deployment Scenario of a NIDS}
\label{ssec:scenario}

\begin{figure}[!t]
    
    \centering
    \includegraphics[width=0.99\columnwidth]{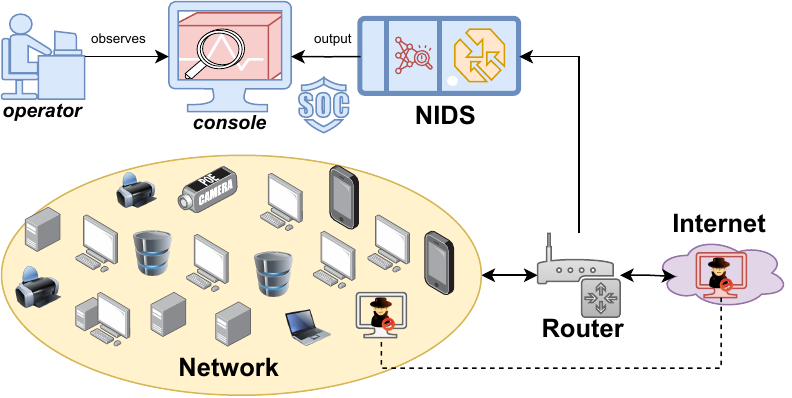}
    \vspace{-3mm}
    \caption{\textbf{Typical NIDS scenario.} 
    \textmd{The NIDS receives data from the \textit{router} as input, and shows its output in a \textit{console}. The NIDS expects intrusions to occur in the network (or from the internet).}} 
    \label{fig:nids}
    \vspace{-2mm}
\end{figure}

\noindent
We provide in Fig.~\ref{fig:nids} a schematic of a typical NIDS deployment in an organization's network. Let us describe Fig.~\ref{fig:nids}.

The network of an organization (yellow oval), contains a variety of heterogeneous devices. This network is connected to the Internet (purple cloud) via a border gateway router---which may potentially integrate some firewall-related network policies~\cite{cropper2015role}. The router forwards all traffic that goes through it to the NIDS, which is positioned on security appliances that are part of the SOC (blue colored objects). The NIDS analyses the network-related data by following any given detection approach. Such network-related data can be in the form of, e.g., raw network-packet captures (i.e., PCAP), or statistical metadata summarizing the network communications occurring in the network (e.g., network flow records, i.e., NetFlow), or a combination thereof~\cite{apruzzese2023sok,Mirsky:Kitsune,barradas2021flowlens,vormayr2020my,cordero2016analyzing,kawanaka2023packet,lu2016hardware,hariri2024rl}.

According to the logic of the detection approach integrated in the NIDS, some alerts will be generated. Such alerts, typically after some post-processing (e.g., alert aggregation or clustering~\cite{van2022deepcase}), are presented on a dedicated \textit{console} (potentially integrated in some SIEM) to the SOC operator---tasked to triage the threat reports that summarize the alerts, and enact the appropriate actions. Unfortunately, in the real world, the majority of these reports tend to be false positives~\cite{alahmadi202299}.
Nevertheless, the figure also shows an attacker that has obtained unauthorized access to a host of the monitored network, which can be controlled remotely (e.g., via a remote-access trojan, or RAT). The NIDS may be able to detect such an intrusion if its detection approach can somehow identify its malicious behaviour (e.g.,~\cite{jiang2015approach}). If this does not happen, the attacker may expand his control on the targeted network (e.g., via lateral movement techniques~\cite{ho2021hopper}), or simply use their infected host as part of a botnet~\cite{idrissi2021toward,zhang2011detecting}, or passively exfiltrate data from such a host~\cite{sabir2021machine}, or do nothing. 

The aforementioned scenario (portrayed in Fig.~\ref{fig:nids}) can be extended to cover also those cases in which, e.g., the NIDS is positioned in a subnetwork of the organization's network, and therefore the NIDS receives its input data from another routing appliance. It can also be positioned on a host that is running various virtual machines---and the NIDS is tasked to monitor the network operations of such virtual machines. Finally, the scenario can also envision a network of a small organization, which may not have a dedicated SOC and either outsources its security to a third-party company/vendor~\cite{apruzzese2022role}; or only has a few IT/network experts that manage its security. 

Given the above, and by applying elementary security principles, we can derive the following considerations:
\begin{itemize}[leftmargin=*, topsep=0pt]
    \item The NIDS, being a security-focused appliance (and being conceivably part of an SOC), is to be deployed in a highly secure location---potentially unreachable from the organization's network. We can expect the NIDS to be heavily protected by tight access-control mechanisms.
    
    \item The NIDS receives data from the router, and is tasked to analyse such data, looking for \textit{intrusions occurring in the network}. In other words, the NIDS trusts the data received by the router (even if attackers attempt evasion). It is beyond the NIDS' scope to ensure the correct operation of the router.
    
    \item The console, being part of the NIDS, is protected by the same defensive layers of the NIDS, and its output is assumed to be observable only by highly-privileged users---such as SOC personnel or IT administrators.
\end{itemize}
These considerations are fundamental to understand our assertions.

\section{Threat Modeling: attacks against NIDS}
\label{sec:assertion_1}

\noindent
The first assertion focuses on a fundamental security aspect of attacks \textit{against} NIDS: the threat model.

In designing (and describing) an attack against a NIDS (i.e., an ``evasion'' attempt), it is crucial to first define the characteristics of the envisioned attacker---i.e., a real person (or group) that is motivated to reach a specific goal by following a given strategy, which depends on the knowledge and capabilities of the attacker w.r.t. the targeted system~\cite{biggio2018wild}.

\begin{cooltextbox}
\textsc{\textbf{Assertion 1}.} 
It is non-sensical to assume a scenario in which a network intrusion detection system (NIDS) is expected to work against an attacker that has compromised such a system. 
\end{cooltextbox}

\subsection{Considerations}
\label{ssec:considerations_1}

\noindent
The first assertion is rooted on a well-known fact: \textit{a compromised security system is, by definition, not reliable}. We provide in Appendix~\ref{app:excerpts} some excerpts, taken from notable prior works~\cite{microsoft2025tcb,whitmore2001method,anderson2001security,reiher2018oss,arpaci2018operating} (some over 40 years old~\cite{thompson1984reflections}), supporting \textsc{Assertion~1}.

Before elucidating the implications of this assertion, let us align such an assertion to the NIDS context. Without loss of generality, let us assume an attacker that, after having obtained control of a host within an organization's network, wants to evade the NIDS protecting such a network.

According to Section~\ref{sec:2}, a NIDS receives network-related data as \textit{input}---which is assumed to be trusted. Therefore, it is straightforward that the device (e.g., the router) that collects/forwards the data that the NIDS shall analyze must not have been compromised. If this is not the case (e.g., if the router sends ``fake'' network-related data to the NIDS), then it would be an overly-optimistic assumption to expect the NIDS to work against an attacker with such powers.

Moreover, after receiving some input data, the NIDS uses a variety of techniques to \textit{analyse} such data. Therefore, it is straightforward that the data-analysis pipeline must not have been compromised. For instance, an attacker who can manipulate the internal operations of the NIDS could just default all communications involving the attacker-controlled hosts to not generate any alert. 

Finally, the NIDS produces an \textit{output} presented in a console. Therefore, the console must not have been compromised, too. For instance, if an attacker had write access to the console, then the attacker could delete the alerts raised by the NIDS; or, if the attacker had read access to the console, then they would be able to immediately discern if an alert is raised about their malicious activities, and change their tactic accordingly---before any action is taken by security operators.

Of course, attackers who have compromised the input data, analytical pipeline, or output console of the NIDS, may be capable of a variety of actions that can enable them to evade the NIDS. What we wrote above are just exemplary (and rather extreme) cases, whose purpose is to demonstrate, logically, that a NIDS becomes unreliable if any of these components has been compromised.

\subsection{Implications and Recommendations}
\label{ssec:implications_1}

\noindent
The principle of \textsc{Assertion 1} is that, when designing an attack against a NIDS (i.e., an evasion attack), the researcher must ensure that the attack (or its implementation) does not implicitly assume a compromise of the NIDS itself.

For instance, by referencing the ``adversarial ML'' domain~\cite{ayub2020model}, wherein attackers try to reach their goals by crafting ``adversarial perturbations'', a perturbation applied in the ``feature space'' (see~\cite{pierazzi2020intriguing}) would represent a violation of \textsc{Assertion 1}, since it would imply that either the NIDS' input data (if, e.g., the NIDS analyses raw PCAP), or the data-processing pipeline (if, e.g., the NIDS analyses NetFlows), has been compromised; this was done in~\cite{apruzzese2020deep}. Similarly, precise manipulations of the ``training dataset'', used to train a given ML model used by the NIDS, could be seen as a compromise of the NIDS analytical pipeline, given that the logic of such an ML model depend on its training phase. Finally, threat models envisioning ``black-box'' attackers that can observe the output of the NIDS to refine their strategies would also imply that the NIDS has been compromised---and is, therefore, unreliable.

To comply with \textsc{Assertion 1}, we propose the following recommendation for future work seeking to study attacks against NIDS (or corresponding defenses):

\vspace{1mm}
\textbox{\textbf{Recommendation 1.} The threat model must assume attackers who can only control the hosts from the Internet, or the hosts within the network that they have compromised---which should not include the router that feeds data to the NIDS, the NIDS itself, or the NIDS output console.}
\vspace{1mm}

\noindent
The generic intuition is to avoid envisioning threat models in which the attacker has, in a sense, ``already won''.
For instance, a valid evasion attempt can entail an attacker that, e.g., fragments the packets of their controlled hosts~\cite{anderson2001security}. This would lead to completely different network-related data (e.g., PCAP) forwarded by the router---but would not imply a compromise of the router.\footnote{Note that it is not trivial to replicate the effects of packet fragmentation by manipulating a pre-collected PCAP trace captured by the router~\cite{apruzzese2024adversarial}.}

Importantly, \textsc{Assertion 1} is not to be intended as a constraint in the creation of novel threat models or, worse, a construct to criticize prior work. On the contrary, the goal is to induce a reflective exercise by the researcher: ``\ul{if the attack's methodology \textit{does} implicitly assume that the NIDS has been compromised, then \textit{how can this be done}?}''
Put differently, the researcher should think of ways that would make the overarching attack not trivial.
Otherwise, the researcher should acknowledge that the reason why the attack succeeds is (also) due to a compromise of the NIDS (or its subelements). In this case, however, the researcher should take into account the added cost to compromise the NIDS (e.g., how is it possible that the attacker manipulated the router to send fake data to the NIDS?), and potentially revise the attacker's goal (e.g., exfiltrate a large amount of sensitive data), or revise the attack's methodology (e.g., if the attacker can freely observe the output of the NIDS console, maybe there are better/cheaper alternatives to reach the same overarching objectives).

Note that assuming that the NIDS has been compromised is not \sout{unrealistic} an impossible scenario. As written in a recent piece: ``The reality is that systems will become compromised or will always have untrusted, partially trusted, or compromised elements''~\cite{rashid2025beyond}. Therefore, there is reason even in studying circumstances that do not follow the aforementioned recommendation. For instance, to design a defense against a (very) powerful attacker.

\section{Evaluation of NIDS: datasets and testbed}
\label{sec:assertion_2}

\noindent
The second assertion focuses on a crucial aspect of most research papers: the experimental evaluation.

The de-facto way to verify whether a given claim holds (e.g., an attacker evades a NIDS) is to carry out an empirical assessment that, by using factual evidence, demonstrates that such a conclusion is ``supported~by~the~data.'' 

\begin{cooltextbox}
\textsc{\textbf{Assertion 2.}} The evaluation testbed must account for the real-world characteristics and implications (for both ``benign'' and ``malicious'' hosts) of the NIDS' deployment scenario \textit{envisioned in the research paper}. 
\end{cooltextbox}

\subsection{Considerations}
\label{ssec:considerations_2}

\noindent
The second assertion is related to the ``dataset'' used to carry out the evaluation of NIDS-related research papers. Given our notion of NIDS (Section~\ref{sec:2}), such an evaluation necessitates a dataset containing network-related data.

The network-security domain has been historically known for being affected by an overall lack of publicly-available datasets that could be used to carry out meaningful experiments. This situation seemed to have improved in the last decade, with the release of various ``benchmark'' datasets that could be used by the research community~\cite{bonninghausen2024introducing}. However, as shown in the recent work by Flood et al.~\cite{flood2024bad}, widely-adopted datasets present significant inconsistencies that, in a sense, undermine the conclusions of papers reliant on them (including, e.g.,~\cite{apruzzese2020deep}). 

The problem, however, lies not in the benchmark datasets per-se---which are, ultimately, just data. Rather, the problem is in the fact that researchers may ``blindly'' rely on datasets containing network-related information that may not align with the scenario envisioned in the threat model. For instance, evaluating a NIDS meant to protect IoT devices on a dataset whose network traffic does not have any IoT device (see~\cite{tagliaro2024large} for a summary) does not make a good argument---regardless of how ``well-crafted'' the dataset is (as also stated in~\cite{ma2023adcl}). At the same time, evaluating a NIDS on a dataset containing malicious traffic generated via methods that were popular decades ago, and showing that the NIDS can detect such attacks, is not a very convincing result. Besides, if the NIDS is expected to be deployed in a ``network of a modern organization'', it would be worrying if a similar organization is not protected against attacks for which known signatures exist.

Nonetheless, an important observation (orthogonal to the aforementioned problem) is that modern networks present immense variability~\cite{Sommer:Outside,apruzzese2023sok}. It is therefore overly-optimistic to expect that a research evaluation, even if carried out over (well-crafted) datasets that capture the behavior of diverse networks, can demonstrate that a given claim holds ``in general'' (i.e., anywhere/anytime).

\subsection{Implications and Recommendations}
\label{ssec:implications_2}

\noindent
It is straightforward that the best way to ensure an evaluation that complies with \textsc{Assertion 2} is to carry out an experiment in a real-world network on which the researcher has (hypothetically) complete control and visibility: this way, the researcher can both {\small \textit{(i)}}~collect data encompassing a variety of ``benign'' activities---ideally pertaining to a large number of heterogeneous hosts; and {\small \textit{(ii)}}~reproduce a given set of ``malicious'' activities---ideally related to recent threats, and used to test the NIDS.

Unfortunately, carrying out such an evaluation may be beyond the reach of most academic researchers (due to, e.g., lack of permission to collect real-users' data, or deploying infected machines in an organization's network). In these cases, we propose the following recommendation:

\vspace{1mm}
\textbox{{\textbf{Recommendation 2:} Ensure that the deployment scenario described in the paper aligns with the network environment captured by the evaluation dataset.}}
\vspace{1mm}

\noindent
The fundamental principle is to \textit{avoid ``overclaiming.''}

There is nothing fundamentally wrong with using publicly available benchmark datasets. However, before carrying out an evaluation on a similar testbed, the researcher should first try to understand what the datasets contain (in terms of both benign and malicious network activities). 

Regardless, relying \textit{only} on benchmark datasets implicitly limits the scope of the evaluation---which can hardly resemble that of a ``modern network'' or ``modern attacks''.
To overcome such a limitation, the researcher can:
{\small \textit{(a)}}~create a dataset in a controlled environment---such as by using physical hardware (as done in, e.g.,~\cite{Mirsky:Kitsune}) or virtual machines (as done in, e.g.,~\cite{dietz2018iot}), or open-source toolkits for network-traffic generation, e.g.,~\cite{cordero2021generating,clausen2019traffic,saez2023gotham}; and/or
{\small \textit{(b)}}~review the most recent network threats, reproduce their behavior (e.g., by leveraging VulnHub~\cite{vulnhub}), and inject them in ``benign'' network activities~\cite{arnaldo2019holy,apruzzese2022cross}---after ensuring that there are no network artifacts which may skew the final results~\cite{arp2022and}. Both of these approaches are: {\small \textit{(i)}}~scientifically valid, {\small \textit{(ii)}}~accessible to any researcher, {\small \textit{(iii)}}~enable some control on the testbed, and {\small \textit{(iv)}}~involve carrying out ``real'' network experiments. Alternatively, a researcher with access to real-world organizational data (which cannot be disclosed) can test a method on such data, and then validate it also on benchmarks: this is what has been done, e.g., in~\cite{talpur2025tagging,ndichu2023machine}.

Finally, and as a corollary to \textsc{Assertion 2}, we advocate that \textit{reviewers} of NIDS papers refrain from overly emphasizing limitations that are inherently insurmountable. For instance, \ul{criticizing a paper because its evaluation cannot prove that a given result holds in general is not a constructive, and almost unfair, remark}---unless the paper itself specifically claims generality. Otherwise, if the paper clearly defines its boundaries, then the crux is verifying if the evaluation's testbed aligns with the deployment scenario described in the paper---and then gauging whether an investigation carried out in such a scenario provides a significant scientific contribution.\footnote{\textbf{Example.} Let us consider the CICIDS17 dataset~\cite{sharafaldin2018toward}, captured in a network of a dozen hosts over the course of one working week. Using such a dataset is valid if it is stated that, e.g., ``we consider a small-scale network of a dozen hosts which generate \scmath{\approx}50GB of network traffic over five days.''} 
\section{Performance Assessment: FPR and TPR}
\label{sec:assertion_3}

\noindent
The third assertion focuses on a specific aspect of the empirical evaluation of an NIDS: the quantitative assessment of the NIDS performance and, in particular, of the false-positive rate~(FPR) and true-positive rate (TPR).

Given that NIDS ultimately seek to ``detect intrusions'', such a task can be seen as a classification problem in which the goal is maximizing the TPR (i.e., correctly classifying ``malicious'' events) while minimizing the FPR (i.e., classifying ``benign'' events as malicious).

\begin{cooltextbox}
\textsc{\textbf{Assertion 3.}} Depending on the specific context, the TPR and the FPR can be incomplete performance metrics: they \ul{not necessarily} are meaningful indicators of a NIDS' practical utility. 
\end{cooltextbox}

\subsection{Considerations}
\label{ssec:considerations_3}

\noindent
The third assertion is strictly related to the base-rate fallacy problem~\cite{axelsson2000base}: modern networks generate such a large amount of data that some false alarms are bound to happen, leading to ``alert fatigue'' that overwhelms security operators, decreasing their efficiency in fulfilling their security duties~\cite{alahmadi202299,mink2023everybody}.

However, from the viewpoint of a SOC operator, it is not necessary to inspect all alerts, e.g., raised by a given NIDS' submodule.
Consider, for instance, the results achieved by Engelen et al.~\cite{engelen2021troubleshooting} on the ``fixed'' version of a well-known dataset containing a mix of benign and malicious NetFlows (collected via simulations lasting around one working week). Specifically, the dataset included \smamath{1\,823\,964} benign and \smamath{881\,211} malicious NetFlows: 75\% of these have been used to train a classifier, which has been tested on the remaining 25\%; hence, the test set included \smamath{455\,991} benign samples, and \smamath{220\,302} malicious samples. Looking at the classification results reported in~\cite{engelen2021troubleshooting}, the classifier correctly detected \smamath{\approx}\smamath{219\,492} malicious samples, and \smamath{\approx}\smamath{451\,431} benign samples,\footnote{These numbers have been obtained by looking at TABLES~I and~II of~\cite{engelen2021troubleshooting}. For TABLE~II, the metrics are provided with only two decimal points, so there may be slight discrepancies in the overall counting. We stress that~\cite{engelen2021troubleshooting}'s contribution was fixing a previously proposed dataset.} meaning an FPR of \smamath{\approx}\smamath{0.01} with a TPR of \smamath{>}\smamath{0.99}. Now, setting aside the TPR and FPR, we argue that \textit{it is inconceivable to think that a human operator can manually triage the over \smamath{220}k samples deemed as malicious by the classifier.}

This is not a matter of the FPR or TPR being low or high. Even if the FPR was \smamath{0}, and the TPR was \smamath{1.0}, no human would triage 220k alerts---or, at least, not one-by-one. Recall (from Section~\ref{ssec:terminology}) that, in the real-world, NIDS are typically part of larger systems (e.g., SIEM) that aggregate the output of various modules into a comprehensive report that enables security analysts to quickly take action. (E.g., see Figs.~\ref{fig:elastic} or~\ref{fig:splunk} in the Appendix~\ref{app:glossary}.) For instance, the recent work by Van Ede et al.~\cite{van2022deepcase} clearly shows that practitioners apply post-processing techniques to cluster the plethora of alerts raised by intrusion-detection modules, resulting in more manageable threat reports that can be manually triaged (to some degree).

Note that the aforementioned problem (which affects, e.g.,~\cite{apruzzese2020deep}) is orthogonal to that of reporting just the FPR without providing the overall number of samples---which can also be source of confusion~\cite{anderson2001security,apruzzese2022role}. For instance, very little can be inferred by reporting that a given classifier achieves an FPR of \smamath{0.01} without mentioning how many datapoints are analysed by such a classifier.

\subsection{Implications and Recommendations}
\label{ssec:implications_3}

\noindent
The crux of \textsc{Assertion~3} is that NIDS produce (a lot of) alerts which \textit{may} be indicators of just a handful of ``intrusions'' affecting one (or more) hosts. 

Hence, we propose the following recommendation---addressed at future work seeking to propose/scrutinize NIDS-related modules (e.g., ML-based classifiers):

\vspace{1mm}
\textbox{{\textbf{Recommendation 3:} the performance of a NIDS-related module should be tied to its practical utility---which requires taking into account also post-processing techniques applied to the output of such a module, and the overarching operational context}.}
\vspace{1mm}

\noindent
The underlying idea of this recommendation is \textit{don't stop.}

Consider a paper whose claimed contribution is a NetFlow classifier. Such a classifier can be integrated in a NIDS by pre-processing the raw PCAP into NetFlows, and then having the classifier analyse such NetFlows (which are deemed to be either benign or malicious). Instead of stopping here and simply measuring the TPR and FPR, the researcher should go one step further, and apply post-processing techniques (potentially proposed by prior work, e.g.,~\cite{van2022deepcase}) to the NetFlows deemed malicious by the classifier. It is at this point that the TPR and FPR should be computed: out of all the ``clusters of threats'' detected by the NIDS \textit{thanks to the classifier}, how many were truly security-noteworthy events? 

For instance, let us re-examine the use-case discussed in Section~\ref{ssec:considerations_3}. From the experiments by Engelen et al.~\cite{engelen2021troubleshooting}, it could have been inferred that the majority of malicious samples entailed the same pair of (compromised) hosts. In this case, even if a classifier, different from the one used in~\cite{engelen2021troubleshooting}, missed some malicious samples (thereby achieving a lower TPR w.r.t.~\cite{engelen2021troubleshooting}), the overall number of raised alarms could still enable the identification of an intrusion (meaning that its overall utility would be the same as that of~\cite{engelen2021troubleshooting}). It could also be that, in a non-critical context, a classifier having, on the surface, a significantly lower TPR (due to missing many malicious NetFlows) could have a higher utility if, e.g., it can detect at least some malicious NetFlows concealing a very subtle intrusion.
At the same time, a classifier raising thousands of ``false positives'' may not be a problem in practice if such alarms are all related to the same host, resulting in a single ``false report'' that would not cause excessive burden to a security analyst.

In short, we advocate that the FPR and TPR are assessed by referring to what the analyst would observe by looking at the security console---and not by focusing solely on a single component of the overarching NIDS.\footnote{Practically, this can be done by implementing some post-processing modules that, e.g., ``if a host generates more than [threshold] malicious samples, then raise an alarm''~\cite{meschini2024case}. Clustering mechanisms can also be employed to aggregate similar datapoints~\cite{van2022deepcase}. In both of these cases, two classifiers having different TPR may have the same utility as they would both raise the same number of alerts.}

\section{Demonstration {\normalsize (Novel Experiments)}}
\label{sec:demonstration}

\noindent
In what follows, we showcase a practical way to use our Assertions in a research context. To this purpose, we carry out novel experiments that leverage well-known and publicly available resources, guaranteeing complete reproducibility of our results (our experimental source code is provided in our repository~\cite{repository}).

\subsection{Setting and Threat Model}
\label{ssec:setting}

\noindent
This subsection focuses on applying \textsc{Assertion~1}.

\textbf{Considered network.} 
We envision a scenario of a small-scale (and non-critical~\cite{knapp2024industrial}) network comprised of a dozen of hosts. Such hosts encompass a variety of operating systems (OS), spanning across both Windows and UNIX-based OSs. These hosts can communicate with hosts from ``external'' networks, which can be reached through a router which also integrates some firewall mechanism. The router, alongside forwarding the traffic to the respective hosts, is also tasked to automatically generate the network flows (NetFlows) of the corresponding traffic.
The NetFlows produced by the router are sent to a classifier that is tasked to infer if they are ``benign'' or ``malicious'' (i.e., a simple binary classification task). 
In particular, the classifier has been trained so that it can discriminate legitimate NetFlows from malicious NetFlows belonging to a variety of well-known attacks, such as:
SSH/FTP-bruteforcing (via \texttt{patator}~\cite{patator}); 
DoS attempts (via \texttt{slowloris}\footnote{https://github.com/gkbrk/slowloris}, 
\texttt{slowhttptest}~\cite{slowhttptest}, 
\texttt{Hulk}~\cite{hulk}, 
\texttt{GoldenEye}~\cite{goldeneye})
; as well as XSS, SQL Injections, the  \texttt{HeartBleed}~\cite{heartbleed} OpenSSL exploit, and some specific Infiltration attempts related to DropBox. Prior work has shown that NetFlows are ideal to detect similar threats~\cite{zhang2012flow}.

\textbf{Security system.} The network is protected by a NIDS. The NIDS receives as input the NetFlows produced by the router, as well as the output of the classifier. These results are shown to a network operator who must make a decision on whether to investigate an internal host that is (potentially) subject of an attack, or not. Due to the non-critical and small-scale nature of the network, human-checks are carried out in batches and on a daily basis.

\textbf{Attacker.} We assume an attacker who gained access to some of the hosts of the network (e.g., via some zero-day exploit, or some unpatched vulnerability~\cite{heo2018knocking}). The attacker wants to cause some damage to this network. The attacker is aware that the network is protected by some NIDS that integrates a classifier trained on samples of the aforementioned attacks. The attacker is not aware of the exact training data, and is also not aware of the specifics of the classifier. The attacker cannot make any query to, e.g., see the output of the classifier. To achieve her objective, the attacker relies on attacks that are not within the training set of the classifier: the DDos \texttt{LOIT}~\cite{loit}, and malware related to the \texttt{ARES} botnet~\cite{ares}.

\vspace{1mm}
\textbox{\textbf{Compliance with \textsc{Assertion~1}.} The aforementioned scenario describes an hypothetical setting that matches the assumptions of NIDS. The attacker has gained a foothold in the network, and the NIDS must detect the attacker's presence. The attacker has compromised only hosts within the network, but has no control on any component within the overarching security system.}

\subsection{Implementation and Pre-Assessment}
\label{ssec:implementation}

\noindent
Here, we show how to comply with \textsc{Assertion~2}.

\textbf{Dataset.} One way to assess the envisioned scenario is by using the well-known CICIDS17 dataset~\cite{sharafaldin2018toward}. Indeed, looking at the description of this dataset, the scenario matches perfectly: the data was captured in a network with \smamath{<20} hosts over the course of five days. On the last day, the creators of CICIDS17 launched attacks conforming to \texttt{LOIT} and \texttt{ARES}, whereas all other attacks are captured on the previous days. Therefore, a good way to practically reproduce our envisioned scenario by using CICIDS17 is to train an ML model on the data of the first four days, and then use the data of the fifth day as test (such data also includes benign traffic, which is fundamental for false positives). Such an experimental protocol follows best practices (i.e., no temporal snooping~\cite{arp2022and,andresini2021insomnia,apruzzese2023sok}).

\begin{table*}[!t]
    \centering
    \caption{\textbf{Case Study Results.} \textmd{We show the results of our novel experiment. We report the TPR, FPR, and total number of errors (misclassifications) achieved by our four considered classifiers when they processed the validation set and the test set at the ``NetFlow-level''; and then show the TPR and TPR when the focus is on identifying internal hosts involved in malicious communications (which we do by post-processing the output of the classifiers).}} 
    \label{tab:performance}
    \vspace{-4mm}
    \resizebox{1.6\columnwidth}{!}{
        \begin{tabular}{c?r|r|r||r|r|r||r|r|r}
             \toprule
             & \multicolumn{3}{c||}{Validation set (NetFlows)} & \multicolumn{3}{c||}{Test set (NetFlows)} & \multicolumn{3}{c}{Test set (Hosts)} \\
             
             Classifier& TPR & FPR & \#Errors & TPR & FPR & \#Errors & TPR & FPR & \#Errors \\
             \midrule

             DT & 0.9996 & 0.0003 & 45 & 0.0636 & 0.0002 & 238,807 & 1.000 & 0.250 & 2 \\
             RF & 0.9996 & \smamath{<}0.0001 & 16 & 0.3706 & 0.0000 & 160,515 & 0.125 & 0 & 7 \\
             HGB & 0.9998 & \smamath{<}0.0001 & 9 & 0.3729 & \smamath{<}0.0001 & 159,913 & 0.125 & 0.125 & 8 \\
             LR & 0.9373 & 0.0711 & 8,068 & 0.3734 & 0.0564 & 164,722 & 1.000 & 1.000 & 8\\
             
             \bottomrule
        \end{tabular}
    }
    \vspace{-2mm}
\end{table*}

\textbf{Development.} We are aware that the CICIDS17 is flawed~\cite{flood2024bad}, so we will use the variant provided by Engelen et al.~\cite{engelen2021troubleshooting}. To further align our setup with prior work, we will use the open-source code in~\cite{apruzzese2023sok}, wherein a complete re-assessment of existing ML-based methods has been carried out---and which also encompasses the CICIDS17 dataset. Practically, we proceed as follows: 
\begin{enumerate}[leftmargin=*, noitemsep,topsep=0pt]
    \item we take the PCAP data of CICIDS17~\cite{sharafaldin2018toward}, extract the NetFlows and label them according to~\cite{engelen2021troubleshooting}, and apply the pre-processing steps of~\cite{apruzzese2023sok}, which distinguish internal-from-external hosts and also standardize the ports according to the IANA groups~\cite{iana}, as well as subsampling data (extracting \smamath{\approx}500k benign NetFlows, and at most 133k malicious NetFlows per attack type).
    
    \item We split the data into train and test: all NetFlows generated on the last day of the capture are considered as test, whereas all the rest is considered as training. Such a split yields \smamath{\approx}342k NetFlows in the test set (of which \smamath{\approx}87k are benign, denoting the large number of malicious NetFlows of the DDoS attack), and \smamath{588}k in the train set (of which \smamath{\approx}409k are benign). 

    \item We train a variety of ML-based classifiers: decision tree (DT), random forest (RF), histogram-gradient boosting (HGB), logistic regression (LR).\footnote{Of course, in practice the NIDS would use only one classifier, but for the sake of this experiment we consider many to show the importance of \textsc{Assertion~3}} All these classifiers support binary classification. Before training each classifier, we split the training set into train:validation with an 80:20 split (note that~\cite{apruzzese2023sok} found no statistically-significant difference in using different types of split on this dataset). The intention is to validate the performance of these classifiers before their deployment: if their performance (on data ``similar'' to that used to train them) is subpar, then such classifiers would not be deployed in practice.
\end{enumerate}
Afterwards, we will test the classifiers on the test set.

\textbf{Preliminary assessment.} Our classifiers perform well on the validation set. 
The DT achieved a TPR of 0.9996 and a FPR of 0.0003 (45 total misclassifications).
The RF achieved a TPR of 0.9996 and a FPR of \smamath{<}0.0001 (16 total misclassifications).
The HGB achieved a TPR of 0.9998 and a FPR of \smamath{<}0.0001 (9 total misclassifications).
The LR achieved a TPR of 0.9373 and a FPR of 0.0711 (8,068 total misclassifications). Note that these results align with those in~\cite{apruzzese2023sok} (even though this specific experiment had not been carried out in~\cite{apruzzese2023sok}). Moreover, we can see that the LR appears to be the worst classifier of the group. The best choice -- by looking at the FPR and TPR -- is the HGB (which is also the fastest to train, only 6s). 

\vspace{1mm}
\textbox{\textbf{Compliance with \textsc{Assertion~2}.} Our experimental setup aligns with our envisioned scenario. We will not attempt to generalize our results: whatever finding we will derive from our experiments, we will not claim that it can be extended to enterprise networks or to IoT scenarios (neither of these are covered by CICIDS17).}

\subsection{Results and Post-processing}
\label{ssec:results}
\noindent
Here, we show how to comply with \textsc{Assertion~3}.

\textbf{TPR and FPR on the test set.}
We assess our classifiers the test set. We can expect that the performance, especially the TPR, will be low: this is because the attacks in the test set stem from generative processes different from those on which the classifiers have been trained. However, one of the promises of ML-based classifiers is to ``detect unseen attacks''~\cite{Sommer:Outside,apruzzese2022role}, so we hope that at least some samples will be detected---especially given that the classifiers are trained on some form of DDoS attack. The results are as follows. 
For the DT, TPR=0.0636 and FPR=0.0002 (238,807 total misclassifications).
For the RF, TPR=0.3706 and FPR=0 (160,515 total misclassifications).
For the HGB, TPR=0.3729 and FPR=\smamath{<}0.0001 (159,913 total misclassifications).
For the LR, TPR=0.3734 and FPR=0.0564 (164,722 total misclassifications).
From these results, we can derive that the ``best'' classifier at test time is either RF or HGB. Indeed, LR has the highest TPR, but the FPR is very high; the DT has the worst TPR of the group; HGB and RF are similar: RF has 0 FPR which is appreciable, but the HGB has also a very low FPR and a slightly higher TPR.

\textbf{A reflection.} Drawing conclusions based on the results above would be misleading. Recall that a NIDS' output is meant to be observed by an operator (§\ref{sec:2}) who would make a choice. If the network operator monitoring this network would see the output of the aforementioned classifiers, what would they do? Even the classifier that raised the lowest number of ``positives'', the DT, predicted that 16,247 NetFlows are malicious. A human analyst cannot manually triage all such NetFlows to make a decision. Indeed, we quote a statement we made in describing our envisioned scenario (emphasis ours): ``These results are shown to a network operator who must make a decision on whether to investigate an \textit{internal host that is (potentially) subject of an attack}, or not.'' Hence, what our envisioned operator would do is check which internal hosts are included in NetFlows that have been flagged as malicious by the classifier.

\textbf{The real performance.} To examine the workflow of our envisioned operator, we developed a simple module that reports how many internal hosts have been involved in the malicious NetFlows flagged by each classifier. For reference, a classifier having accuracy of 100\% would flag that 8 internal hosts should be investigated (whereas the remaining 8 hosts that made some communications on this day should not). The results are as follows.
\begin{itemize}[leftmargin=*, topsep=0pt]
    \item The DT flagged 10 hosts: 8 are involved in malicious NetFlows (i.e., the ``true positives''), whereas 2 are not (i.e., ``false positives''). 
    \item The RF flagged a single host, which was involved in malicious communications. However, it missed all the other 7 hosts.
    \item The HGB flagged two hosts: one was indeed involved in malicious communications, the other was not.
    \item The LR flagged 16 hosts: 8 were indeed involved in malicious communications, but 8 were not. 
\end{itemize}
So, in practice, if the operator were observing the output of the DT, then the TPR would be 1.000 (because 8 out of 8 hosts-that-should-be-investigated would be truly investigated) and the FPR 0.250 (because the operator would inspect two hosts that did not have any malicious activity). Conversely, if the operator relied on the RF, then the TPR would be 0.125 and the FPR=0; and for the HGB, the TPR would also be 0.125, but the FPR=0.125. Finally, for the LR, both TPR and FPR would be 1. From this perspective, the DT appears to be much better than both the HGB and the RF---even though the performance of the DT was statistically-significantly (\smamath{p}<.05) lower than both of them on the test set (from the viewpoint of individual NetFlows classification) and also lower in the validation set. 

\vspace{1mm}
\textbox{\textbf{Compliance with \textsc{Assertion~3}.} Instead of stopping at the results of individual classifiers, we went one step further and examined how the classifier's output could be integrated in an operational context. By adopting a different perspective and post-processing the output, we found that a classifier that appeared to be suboptimal (classification-wise) may be the best (utility-wise).}
\vspace{1mm}

We conclude with a remark. We do not claim that our results hold in general. And we also do not claim that DT-based classifiers are always better than RF-/HGB-based ones. We also do not claim that our post-processing module is the one adopted in the real-world, nor that security operators would behave exactly in the same way we envisioned in our case study. We reiterate that this is a SoK paper meant for researchers. We provide the complete results in Table~\ref{tab:performance} (the code in our open-source repository~\cite{repository}).

\section{Discussion and Related Work}
\label{sec:discussion}

\noindent
We critically examine the main contributions of this work, i.e., the assertions and their recommendations~(§\ref{ssec:critical}); we also expand our discussion with practical observations~(§\ref{ssec:observations}).
Then, we provide some anecdotes serving as a motivation for our work~(§\ref{ssec:motivation}). In doing so, we also cover some related works. Finally, we provide our vademecum and takeaways (§\ref{ssec:takeaways}).

\subsection{Critical Remarks {\normalsize (Frequently Asked Questions)}}
\label{ssec:critical}

\noindent
To avoid generating misunderstandings, we discuss potential counterarguments to the \textsc{assertions} made in this paper. We do so by adopting a question and answer format.

``\textit{With regards to `adversarial perturbations', wouldn't \textsc{Assertion 1} imply that studying any form of adversarial ML attack is pointless in the NIDS context?}''
No. As stated in §\ref{ssec:considerations_1}, \textsc{Assertion 1} is not meant to be ``a constraint in the creation of novel threat models or, worse, a construct to criticize prior work.'' Although \textit{some} adversarial ML attacks envisioned by prior work (e.g.,~\cite{apruzzese2020deep}) may have entailed ``unrealizable'' perturbations---potentially stemming from attackers that had (implicitly) already ``breached'' the NIDS---this does not undermine the value of any such prior (or even future) publication. 
\begin{itemize}[leftmargin=*,topsep=0pt]
    \item First, because it is not wrong in a \textit{research paper} to make assumptions that do not fully align with the real world. 
    \item Second, because any ``misalignment'' may not be apparent at the time of publication (after all, research papers are at the forefront of human knowledge), and can be used as a lesson learned for future research (as an example:~\cite{arp2022and}). 
    \item Third, because even an \sout{unrealistic} unlikely assumption (e.g., an attacker who has compromised the overarching security system) may still occur in reality. 
    \item Fourth, because even if the likelihood of such an occurrence is low, from a research viewpoint it is still insightful to study some security/robustness properties of any given NIDS component to investigate worst-case scenarios. 
\end{itemize}
Nevertheless, we stress that it is possible to stage adversarial ML attacks against NIDS that do not assume a compromise of the overarching security system. For instance, attacks at \textit{training-time} (so-called ``poisoning'') can be launched just from a single attacker-controlled host---we strongly recommend looking into the threat model envisioned in the recent work by Severi et al.~\cite{severi2023poisoning}; whereas attacks involving adversarial perturbations crafted at \textit{inference-time} require a careful treatment of the ``problem space''\footnote{For instance, it is crucial to identify how a real-world attacker can, from their controlled hosts, perform actions that affect the feature representation of a sample that is ultimately analysed by (and then evades) an ML model.} but are certainly plausible. For noteworthy examples, we point the reader to the recent papers by Erba et al.~\cite{erba2024practical} and Catillo et al.~\cite{catillo2024towards}, both of which discussing ways attackers can use to influence the data seen by ML models---while complying with \textsc{Assertion 1}. 

``\textit{Many works have examined prior literature (on various forms of intrusion detection) under a real-world lens. What is the added value of \textsc{Assertion 2}, then?}''
It is true that many papers (e.g.,~\cite{pendlebury2019tesseract,arp2022and,apruzzese2021modeling,apruzzese2023sok,catillo2023machine,flood2024bad,ceschin2024machine}) have, explicitly or implicitly, scrutinized the testbed or evaluation methodology of related works from a realistic viewpoint. However, the takeaways of all of these contributions are orthogonal to that of \textsc{Assertion 2}. Specifically, we focus the attention to the recommendation written in §\ref{ssec:implications_2}: we argue that, in principle, it is scientifically acceptable for a research paper to carry out an evaluation in a ``toy'' testbed---as long as the paper does not claim that the results hold in the real world. 
However, we also argue that given the impossibility of replicating the behavior of ``all'' real-world networks (or to test the effectiveness of an NIDS ``in general''), reviewers should lower their expectations---and not use the fact that a paper carries out an evaluation via a (potentially well-done) simulation as a ``weakness'' of a paper just because ``simulations cannot replicate the real world''. We are not aware of any work (in the intrusion detection context) that advocated for such recommendations---which, we stress, are addressed to reviewers, too (and hence pertain to ``metascience'' research).\footnote{An argument can be made about the NIDS vs malware-detection domains in peer-review contexts. For the latter, we believe that research is more aligned with the real world (e.g.,~\cite{kaya2025ml,ceschin2019shallow}), because real-world malware and goodware examples are easier to find and use for research. In contrast, NIDS-related data is closed-source, and experiments require simulations which are easy to criticize (our view is in §\ref{ssec:implications_2}).}

``\textit{For \textsc{Assertion 3}, wouldn't neglecting the FPR/TPR prevent one from measuring the value of a specific NIDS submodule (which can be a paper's main contribution)?}''
This is a valid remark---which is, however, orthogonal to the point of \textsc{Assertion 3}. Specifically, the point is not to ``neglect'' the TPR or FPR (which \textit{can} be appropriate performance metrics); rather, the point is that, in some cases, it may be more appropriate to focus on the real-world utility of a given NIDS submodule (e.g., a classifier). Such utility can only be measured by taking into account the overall system performance. At the same time, and from the viewpoint of a peer-reviewer, it may be unfair to criticize the FPR (or TPR) for being ``too high'' (or ``too low''): such critiques should be done by accounting for the role played by these metrics in the overarching NIDS---and not by merely looking at the results of a single classifier (in our experiment, the TPR of the DT was the worst on the NetFlows, but near-perfect when identifying hosts involved in malicious communications---see Table~\ref{tab:performance}.) Rather than critizing the low/high TPR/FPR, a reviewer may suggest to discuss how the FPR/TPR of a classifier can impact the system-wide effectiveness of the envisioned NIDS. 

``\textit{Must any future work follow/comply with the three \textsc{assertions} stated in this paper to provide a meaningful contribution to the state of the art?}''
No. The scientific contribution of any research paper (past or future) on NIDS is independent from how well it embraces any of the assertions proposed in our work. Ultimately, ours is a reflective piece: researchers are free to disagree with the statements made in this work. However, we also believe that the assertions (and corresponding recommendations) proposed in this piece provide some original viewpoints---addressed also to researchers serving as peer-reviewers---that can be beneficial for this research domain (such as, e.g., not recommending to ``reject'' a paper because there is no evaluation on a real enterprise network).\footnote{Reviewers should be more mindful of how NIDS relate to the real world.}

\subsection{Practical Observations}
\label{ssec:observations}
\noindent
Here, we provide some complementary observations, rooted on established prior work and operational best practices, that expand the contributions of this piece. The intention is to inspire future developments in the NIDS domain from a research viewpoint.\footnote{These observations have been derived during the peer-review phase of this paper.}

\textbf{Attacker's Knowledge.} \textsc{Assertion 1} predominantly focuses on the \textit{capabilities} of the attacker. However, the \textit{knowledge} of the attacker, which is a crucial component of an adversarial threat model, is orthogonal to the capabilities~\cite{apruzzese2023real}. For instance, an attacker can have ``perfect knowledge'' of a system without having compromised such a system: this can happen if, e.g., an open-source implementation of such a system is available and a company uses such a prototype in production (see, e.g.,~\cite{nasr2025evaluating}). However, and as remarked in~\cite{apruzzese2021modeling}, having complete knowledge of the entire system is challenging from the viewpoint of a real-world attacker, since it would require insider's knowledge---which is feasible, but costly. According to~\cite{apruzzese2021modeling}, the most likely use-case of attackers trying to bypass an NIDS is that of a ``partial-knowledge adversary'', who may reasonably infer (potentially via OSINT) some information about the NIDS (or the underlying network environment) used by a targeted company. In summary, we believe that making sensible assumptions on the attacker's knowledge to be important, but complementary to \textsc{Assertion 1}.

\textbf{Approximating Real-world Scenarios.} As we stated for \textsc{Assertion 2} (§\ref{ssec:considerations_2}), some researchers may not have access to data from real-world networks for experiments. This, however, should not prevent one from carrying out evaluations that \textit{resemble} real-world scenarios. We believe that, for research papers, even a proof-of-concept (but well-executed) simulation to be a valid way to assess certain claims (e.g.,~\cite{csikor2021privacy}).
Especially recently, many works proposed tools for network-traffic simulations (e.g.,~\cite{verkerken2026concap, uetz2021reproducible, wang2022tool, saez2023gotham}) usable to mimic real-world setups---potentially by enriching existing benchmark datasets with more recent datapoints~\cite{verkerken2026concap}. Nonetheless, sharing the entire artifacts (including, e.g., documentation on how the simulation was carried out---and not just the data itself) can substantially improve the quality of an experimental evaluation, since downstream research can determine how well the testbed used in prior work can be used as a basis for new experiments.

\textbf{Detection Timeliness.} In some contexts (e.g., critical infrastructures), alerts should be triaged immediately, and malicious communications should be found as early as possible. In these cases, a classifier with perfect TPR (assuming low FPR) is desirable because it would detect ``all'' malicious attempts---which may lead to an earlier response by human operators (who should triage such alarms in near-real time). However, the sheer number of events generated in modern infrastructures is massive~\cite{tariq2025alert}, making it unfeasible to triage every single alert (e.g., the authors of DeepCase~\cite{van2022deepcase} consider ``throttling'' periods varying between 15-minutes and 1 day). Nevertheless, we therefore solicit researchers to \textit{clearly state how important the timeliness of the detection is for the envisioned scenario} (to support \textsc{Assertion 3}). For instance, in our demonstration (§\ref{sec:demonstration}), we explicitly stated that the human analysts triages alarms on a daily basis---which is a sensible assumption given that we envisioned a small-scale and non-critical network.

\textbf{Usages of a Classifier's Output.}
A corollary of \textsc{Assertion 3} is that the output of a classifier (e.g., whether an input is benign or malicious) can be itself processed by additional models and methods (see Section~\ref{ssec:implications_3}). Our demonstrative experiment (in Section~\ref{ssec:results}) showed a simple use case of ``aggregator of host-specific detections'', which was meant to be provided to a human analyst for triaging. However, the pipeline between classifier output and human decision can be made up of various components that can cross-correlate information derived from various sources and thus reduce the number of alerts that must be manually triaged~\cite{jalalvand2024alert, baruwal2024towards}. For instance, a recent work~\cite{jaworski2026hybrid} combines alerts with threat-intelligence; whereas~\cite{bryant2020improving} combine alerts from multiple sensors; while~\cite{roelofs2024finding} combine outputs of multiple defensive schemes. Put simply, there are numerous ways in which the output of a given component (such as a NetFlow classifier) can be integrated in an operational security system. Hence, the quality of a component should not be solely assessed in isolation, but by accounting for the overarching system's performance and the respective deployment context. From the viewpoint of a researcher (and to echo our Recommendation-3 in Section~\ref{ssec:implications_3}), we thus endorse exploring all these possibilities---which would be facilitated by disclosure of open-source prototypes.

\subsection{Motivation}
\label{ssec:motivation}

\noindent
As stated in the Introduction (§\ref{ssec:goal}), the contributions of this work are built on the assumption that ``research in NIDS-related areas is relatively stagnant, and is hindered by---among others---a superficial treatment of the NIDS domain.'' 

In what follows, we provide factual evidence suggesting that such an assumption holds (as of September 2025). In doing so, we maintain the same rationale followed in the rest of the paper: we \textit{are not criticizing any prior work}.

\begin{figure}[!t]
    \centering
    \includegraphics[width=0.99\columnwidth]{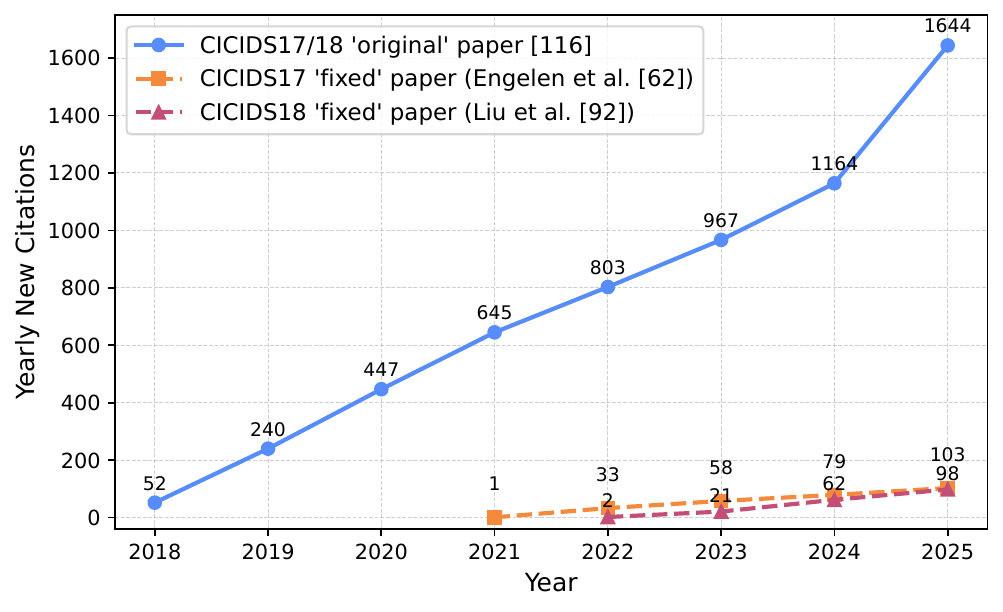}
    \vspace{-4mm}
    \caption{\textbf{Comparison between~\cite{sharafaldin2018toward} and~\cite{engelen2021troubleshooting, liu2022error}.} \textmd{We plot the new citations achieved by~\cite{sharafaldin2018toward} (original paper of CICIDS17/18) and the work by Engelen et al.~\cite{engelen2021troubleshooting} (paper that ``fixed'' CICIDS17) and Liu et al.~\cite{liu2022error} (paper that ``fixed'' CICIDS18) every year since their publication (Source: Google Scholar).}}
    \label{fig:comparison}
    \vspace{-3mm}
\end{figure}

\begin{itemize}[leftmargin=*, topsep=0pt]
    \item During the interactive author-reviewer discussion\footnote{Publicly observable at: \url{https://openreview.net/forum?id=zfCNwRQ569}} of a NeurIPS'23 paper on NIDS~\cite{li2023interpreting}, a reviewer pointed out that not enough information was provided on the features used to train the ML models. The authors responded that ``\textit{Considering that NeurIPS focuses on AI, we didn’t include too much details on network traffic in our paper.}'' Such details were later added in the final version of the paper.~\smamath{\Rightarrow}~While this is an example of a constructive peer-review, this anecdote suggests that in some (top-tier and ML-centered) research venues, the NIDS domain may be treated superficially.
    
    \item Many prior works (e.g.,~\cite{alahmadi202299,apruzzese2023sok,mink2023everybody,alahmadi202299}) have explicitly pointed out the skepticism, or uncertainty, of practitioners towards techniques proposed in NIDS-related research. And this is surprising given the thousands of publications on this subject (see Fig.~\ref{fig:scholar}). For instance, false alarms were well-known even \smamath{>}25 years ago (see, e.g.,~\cite{axelsson2000base,debar2001aggregation}) and are still widespread today (see, e.g.,~\cite{alahmadi202299,mink2023everybody}). However, the concept of ``alarm'' (or of ``alert'') may not be among the priorities of some NIDS-related researches (e.g., in~\cite{apruzzese2020deep}, the terms ``alarm'' or ``alert'' or even ``report'' are \textit{never mentioned}\footnote{For ``report'', we are referring to its usage as a noun (there are plenty of occurrences of ``report'' as a verb in~\cite{apruzzese2020deep}). In contrast, in~\cite{apruzzese2020deep} there is mention of ``false positive'' which, as argued in \textsc{Assertion 3}, it can be a misleading (or incomplete) metric.}), despite being intrinsic of NIDS (see §\ref{ssec:terminology}).~\smamath{\Rightarrow}~These anecdotes further show that {\small \textit{(i)}}~some fundamental characteristics of NIDS may be overlooked by NIDS-related papers, and {\small \textit{(ii)}}~from a practical viewpoint, research advances in this domain are lackluster.

    \item One of the most popular benchmark datasets for NIDS is CICIDS17, as well as its extension the CICIDS18: according to their creators (see~\cite{cicids18, cicids17}), both of these datasets refer to the same research paper~\cite{sharafaldin2018toward} which counted \smamath{5\,962} citations on Google Scholar since its publication in 2018 and until the end of 2025. However, in 2021, a work by Engelen et al.~\cite{engelen2021troubleshooting} (published in the Workshop on Traffic Measurements for Cybersecurity, co-located with the IEEE Symposium on Security and Privacy) carried out a thorough analysis of CICIDS17 and {\small \textit{(i)}}~found substantial issues while also {\small \textit{(ii)}}~releasing the code to ``fix'' such issues. In 2022, Liu et al.~\cite{liu2022error} found and fixed errors in CICIDS18 (in a work that won the Best Paper Award at IEEE CNS). To provide additional evidence that \textit{some} research papers may overlook the real-world implications of NIDS-related data, we study the ``citation trend'' of these three works, visually shown in Fig.~\ref{fig:comparison}. Indeed, it is reasonable to assume that {\small \textit{(a)}}~a paper using CICIDS17/18 would cite~\cite{sharafaldin2018toward}; and {\small \textit{(b)}}~researchers \textit{aware} of the issues of CICIDS17/18 that would still use this dataset for their experiments would cite~\cite{engelen2021troubleshooting} or~\cite{liu2022error}.\footnote{Indeed, if a paper used CICIDS17 (or CICIDS18)  but does not cite~\cite{engelen2021troubleshooting} (or~\cite{liu2022error}), there is a high risk that the evaluation is carried out on the ``flawed'' data of CICIDS17 (or CICIDS18) without even acknowledging potential threat to validity.} Therefore, in a perfect world, the citation trend of~\cite{engelen2021troubleshooting, liu2022error} and~\cite{sharafaldin2018toward} would align.\footnote{There are other works that discussed issues in CICIDS17 or CICIDS18, such as~\cite{lanvin2022errors,flood2024bad}, but they received even less citations than ~\cite{engelen2021troubleshooting,liu2022error}, hence we omit them from this analysis as our conclusions would not be affected.} However, we can see from Fig.~\ref{fig:comparison} that this is not the case: 
    since 2022 and until the end of 2025, the works by Engelen et al.~\cite{engelen2021troubleshooting} and Liu et al.~\cite{liu2022error} cumulatively accrued \smamath{457} citations, whereas the publication of CICIDS17/18~\cite{sharafaldin2018toward} obtained \smamath{4\,578} citations during the same timespan.~\smamath{\Rightarrow}~This result shows that prior work may overlook the issues affecting the real-world validity of the data used to test a given NIDS-related hypothesis.\footnote{Note that it would be unfair to question the ecological validity of papers (such as~\cite{soltani2023adaptable}) considering CICIDS17 but which have been published before~\cite{engelen2021troubleshooting}.} 

    \item Expanding on the point above, even in top-tier conferences both {\small \textit{(i)}}~authors and {\small \textit{(ii)}}~program-committee members may not be aware of the issues of CICIDS17 (which have been known since 2021). For instance, a recent IEEE S\&P'24 paper (i.e.,~\cite{diallo2024sabre}) carried out an evaluation on CICIDS17 but without accounting for the findings of Engelen et al.~\cite{engelen2021troubleshooting} (which was presented at a workshop of IEEE S\&P!). Of course, and as a disclaimer, this does not mean that any conclusion based on the ``original'' CICIDS17 is unfounded. For instance, let us consider the author-reviewer discussion\footnote{Publicly observable at: \url{https://openreview.net/forum?id=KYHVBsEHuC}} of a NeurIPS'24 paper on NIDS~\cite{bin2024diffupac}. It was pointed out that the initial submission carried out its evaluation on the original CICIDS17: the authors promptly (and remarkably) re-did their experiments on the ``fixed'' version of CICIDS17, and found that the results still supported the conclusions.~\smamath{\Rightarrow}~Put simply, the point is that there is nothing fundamentally wrong in using the ``original'' CICIDS17 for NIDS-related research; however, focusing excessively on the ``research'' side of a publication (without, e.g., questioning what is contained in a given benchmark dataset) may hinder some real-world implications of its conclusions.

\end{itemize}
To conclude, we believe that all of the above is evidence of some fundamental misunderstandings of the NIDS domain---which may (perhaps unconsciously) affect both authors of research papers, as well as peer-reviewers. As we wrote (see §\ref{ssec:audience}), our paper is addressed at both of these categories of readers. Authors and reviewers can look at some of the arguments made here and use them to reflect on their statements (either in their paper, or in the review).

\newcommand{\tabelement}[1]{{\normalsize #1}}

\begin{table*}[!t]
    \centering
    \caption{\textbf{Vademecum.} \textmd{Questions to ask when approaching NIDS research (Ref. points to this SoK's section). Some questions may not be universally applicable.}} 
    \label{tab:vademecum}
    \vspace{-4mm}
    \resizebox{1.99\columnwidth}{!}{
        \begin{tabular}{c|l|c}
             \toprule
             \textsc{\#} &~~\textsc{Question} & Ref.\\
             \midrule

             Q1 & \tabelement{Has the notion of ``anomaly'' been properly defined?} & {\normalsize §\ref{ssec:terminology}} \\ 
             Q2 & \tabelement{Has the term ``unrealistic'' been used unambiguously?} & {\normalsize §\ref{ssec:terminology}} \\ 
             Q3 & \tabelement{Does the evasion attack's methodology \textit{implicitly} assume a compromise of the NIDS (or of its input/output)?} & {\normalsize §\ref{sec:assertion_1}} \\ 
             Q4 & \tabelement{(if yes to Q3) Has the fact that the NIDS is assumed to be compromised been taken into account, or at least justified?} & {\normalsize §\ref{ssec:implications_1}} \\ 
             Q5 & \tabelement{Does the evaluation dataset capture the characteristics of the network environment envisioned in the threat model?} & {\normalsize §\ref{sec:assertion_2}} \\ 
             Q6 & \tabelement{Does the paper claim ``generality'' of the conclusions---and, if so, is there sufficient evidence to substantiate such a claim?} & {\normalsize §\ref{ssec:implications_2}} \\ 
             Q7 & \tabelement{Has the TPR/FPR been operationally contextualized (e.g., how does the analyst use the output of a classifier)?} 
             & {\normalsize §\ref{ssec:implications_3}} \\ 
             
             \bottomrule
        \end{tabular}
    }
    \vspace{-2mm}
\end{table*}

\subsection{Takeaways and Vademecum}
\label{ssec:takeaways}

\noindent
We have touched a variety of topics pertaining to NIDS---which we endorse researchers to take into account whenever they engage with this domain. To facilitate this, we coalesce our assertions and recommendations into a set of seven questions, shown in Table~\ref{tab:vademecum}. These questions represent the \textit{vademecum} mentioned in the Introduction~(§\ref{ssec:goal}). 
Let us briefly discuss each question (Q).
\begin{itemize}[leftmargin=*, topsep=0pt]
    \item Q1 and Q2 do not refer to a specific \text{Assertion}, but are fundamental to avoid triggering misunderstandings.
    \item Q3 and Q4, related to \textsc{Assertion 1}, are particularly useful for adversarial ML papers.
    \item Q5 and Q6, related to \textsc{Assertion 2}, serve to keep the paper's findings safely anchored on its testbed (which should be carefully examined, especially if publicly-available data is used).
    \item Q7, related to \textsc{Assertion 3}, serves as a reminder to try adopting the viewpoint of a security operator, i.e., the real user of a NIDS.
\end{itemize}
Ideally, these questions can be useful to both {\small \textit{(i)}}~investigators/authors of papers, or {\small \textit{(ii)}}~reviewers. The former can use these questions either at the beginning of a given research project (e.g., to shape a given research plan), or during the course of the research activities, as well as during the writing of an article (e.g., as ``sanity checks''). The latter can use these questions to guide the reviewing process of a submission. Potentially, these questions can expedite the author-reviewer interactions, or help clarifying some doubts.\footnote{We would be delighted if the authors of another submission under review can ``convince'' a referee (e.g., during a rebuttal) by referencing this work!}

\textbox{In our demonstration (§\ref{sec:demonstration}), we followed our vademecum: ``anomaly'' and ``unrealistic'' have never been used; the attacker is not assumed to have compromised the NIDS; the dataset was carefully chosen so as to resemble the envisioned scenario; there is no ``generality claim''; and the TPR/FPR have been assessed by considering the post-processed output of a classifier.}

\section{Conclusions and Future Outlook}
\label{sec:conclusions}

\noindent
We seek to induce a change in the landscape of NIDS research.

The assertions stated in this work are rooted in established security principles and may appear well-known to some readers; similarly, some recommendations also echo those made in security-focused literature. For instance, the seminal work by Sommer and Paxson~\cite{Sommer:Outside}, published over 15 years ago, emphasized, e.g., the necessity of realistic threat models in the NIDS research domain. However, we believe that the takeaways of~\cite{Sommer:Outside} may have faded. Our assertions and vademecum, supported by our original experiment, hence reinforces the messages broadcast by~\cite{Sommer:Outside}.

We hope that the research community on NIDS can be inspired by the arguments and reflective exercises provided in this SoK.

\begin{acks}
    The author would like to thank all the reviewers who provided feedback to previous versions of this manuscript. Parts of this research was funded by Hilti.
\end{acks}

\bibliographystyle{ACM-Reference-Format}


\appendices

\section{Glossary and Images}
\label{app:glossary}
\noindent
We provide a list of technical definitions, taken verbatim from the RFC~\cite{shirey2007internet}, which are crucial for this paper.

\begin{figure}[!t]
    \centering
    \frame{\includegraphics[width=0.99\columnwidth]{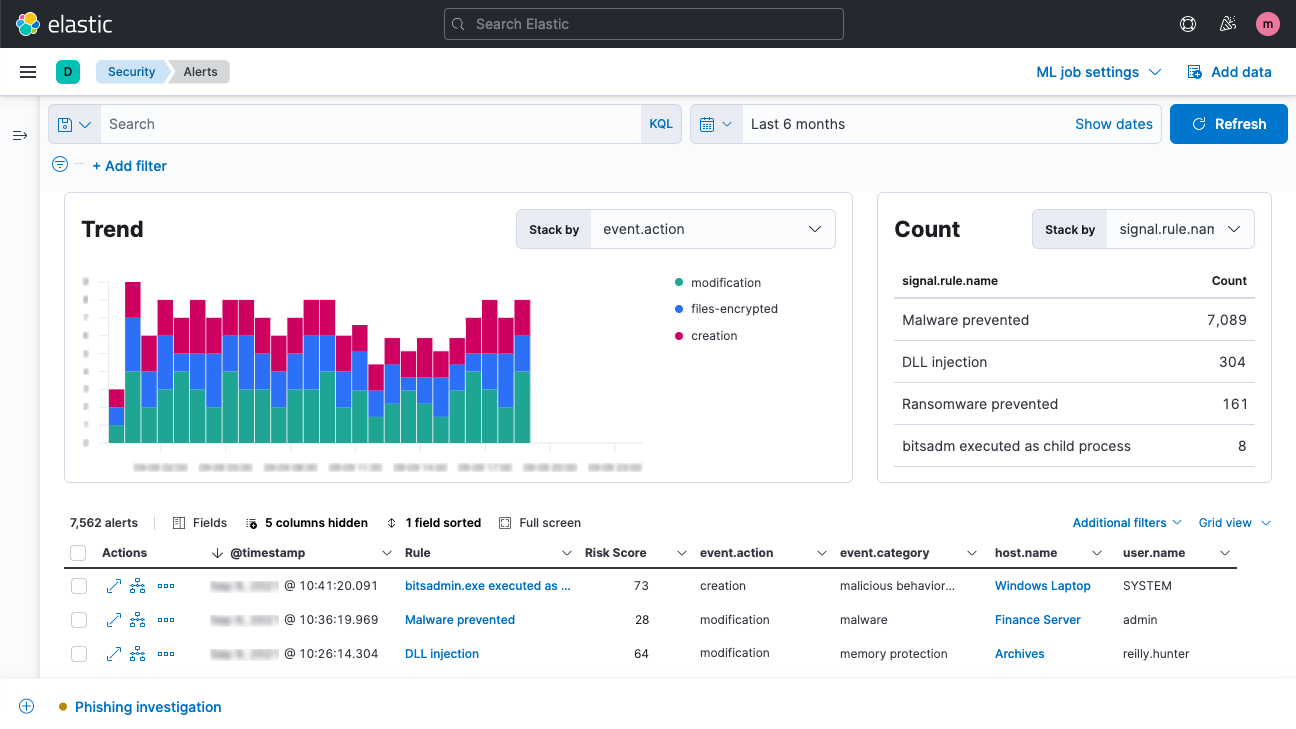}}
    \vspace{-2mm}
    \caption{\textbf{The dashboard of Elastic} --
    \textmd{\footnotesize source:~\cite{elasticsiem}}} 
    \label{fig:elastic}
    \vspace{-5mm}
\end{figure}

\begin{figure}[!t]
    \centering
    \frame{\includegraphics[width=0.99\columnwidth]{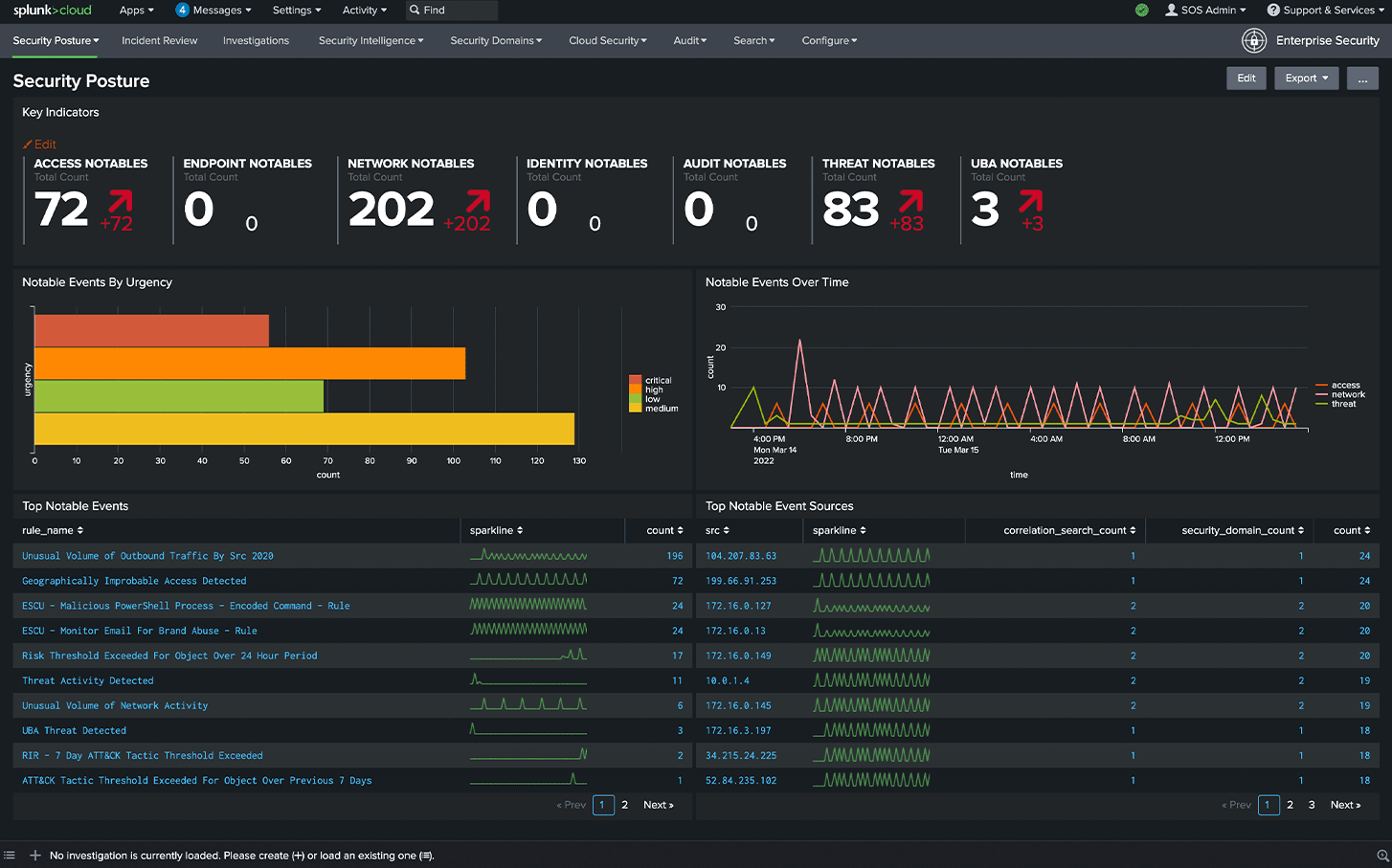}}
    \vspace{-2mm}
    \caption{\textbf{The dashboard of Splunk} --
    \textmd{\footnotesize source:~\cite{splunksiem}}} 
    \label{fig:splunk}
\end{figure}

\begin{figure}[!t]
    
    \centering
    \frame{\includegraphics[width=0.99\columnwidth]{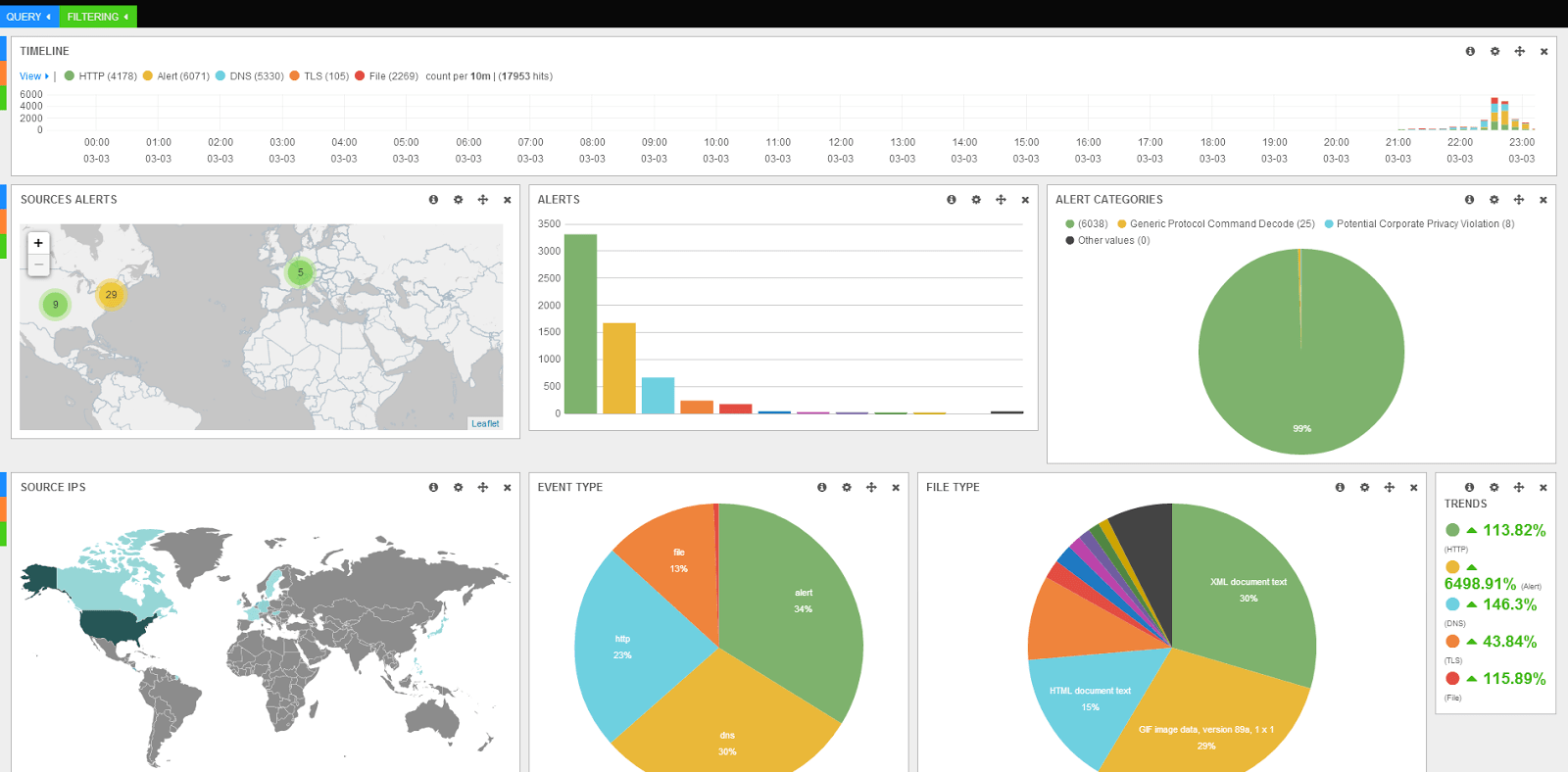}}
    \vspace{-2mm}
    \caption{\textbf{The dashboard of Suricata} --
    \textmd{\footnotesize source:~\cite{suricatasiem}}} 
    \label{fig:suricata}
    \vspace{-5mm}
\end{figure}

\begin{itemize}[leftmargin=*, topsep=0pt]

    \item \textit{Attack}: An intentional act by which an entity attempts to evade security services and violate the security policy of a system.      That is, an actual assault on system security that derives from an       intelligent threat.
      
    \item \textit{(Computer) Network}: A collection of host computers together with the subnetwork or internetwork through which they can exchange data. {\small This definition is intended to cover systems of all sizes and types, ranging from the complex Internet to a simple system composed of a personal computer dialing in as a remote terminal of another computer.}
    
    \item \textit{(Information) System}: An organized assembly of computing and communication resources and procedures -- i.e., equipment and services, together with their supporting infrastructure, facilities, and personnel -- that create, collect, record, process, store, transport, retrieve, display, disseminate, control, or dispose of information to accomplish a specified set of functions.

    \item \textit{Intrusion}: A security event, or a combination of multiple security events, that constitutes a security incident in which an intruder  gains, or attempts to gain, access to a system or system resource without having authorization to do so. (alternate)  A type of threat action whereby an unauthorized entity gains access to sensitive data by circumventing a system’s security protections.
    
    \item \textit{Intrusion Detection System}: A process or subsystem, implemented in software or hardware, that automates the tasks of {\small \textit{(a)}}~monitoring events that occur in a computer network and {\small \textit{(b)}}~analyzing them for signs of security problems. (alternate) A security alarm system to detect unauthorized entry.

    \item \textit{Security Compromise}: A security violation in which a system resource is exposed, or is potentially exposed, to unauthorized access.
    
\end{itemize}
We provide in Figs.~\ref{fig:elastic},~\ref{fig:splunk}, and~\ref{fig:suricata} some exemplary dashboards of state-of-the-art NIDS software (i.e., SIEM).

\section{Excerpts supporting \textsc{Assertion 1}}
\label{app:excerpts}

\noindent
Many seminal works support \textsc{Assertion 1}, and many software houses leverage its underlying a principle to design their solutions. Below are some excerpts (taken verbatim, emphasis ours) that justify \textsc{Assertion 1}.

\begin{itemize}[leftmargin=*, topsep=0pt]
    \item ``The Trusted Computing Base (TCB) refers to all of a system's hardware, firmware, and software components that provide a secure environment. The components inside the TCB are considered `critical.' \textit{If one component inside the TCB is compromised, the entire system's security may be jeopardized}.''~\cite{microsoft2025tcb}
    \item ``No amount of source-level verification or scrutiny will protect you from using untrusted code.''~\cite{thompson1984reflections}.
    \item ``A security audit subsystem is responsible for capturing, analyzing, reporting, archiving, and retrieving records of events and conditions within a computing solution. [...] can include [...] intrusion detection components. Security requirements for an audit subsystem would include: \textit{trusted transfer of audit data}; \textit{protection of security audit data}; analysis of security audit data---including review, anomaly detection, violation analysis, and attack analysis using simple or complex heuristics; alarms for loss thresholds.'' Additionally, the security audit may be used by the ``solution integrity subsystem'' which should focus on (among others) ``\textit{functional isolation using domain separation} or a reference monitor''~\cite{whitmore2001method}.
    \item ``The reference validation mechanism must be Tamper-proof. Without this property, an attacker can undermine the mechanism itself and hence violate the security policy.''~\cite{anderson2001security}
    \item ``As a rule, if the software you are running on top of, whether it be an operating system, a piece of middleware, or something else, is insecure, what’s above it is going to also be insecure''~\cite{reiher2018oss,arpaci2018operating}.

\end{itemize}

\noindent
Altogether, the aforementioned statements support the thesis that, to fulfill its security-related functions, all components of a security-focused system must not have been compromised---which is the crux of \textsc{Assertion 1}.

\end{document}